\documentclass[RNAAS]{aastex62}

\begin{document}

\title{ASAS-SN Identification of FY Sct as a detached eclipsing binary system with a $\sim2.6$ year period}

\correspondingauthor{T. Jayasinghe}
\email{jayasinghearachchilage.1@osu.edu}

\author[0000-0002-6244-477X]{T. Jayasinghe}
\affiliation{Department of Astronomy, The Ohio State University, 140 West 18th Avenue, Columbus, OH 43210, USA}
\author{K. Z.  Stanek}
\affiliation{Department of Astronomy, The Ohio State University, 140 West 18th Avenue, Columbus, OH 43210, USA}
\affiliation{Center for Cosmology and Astroparticle Physics, The Ohio
State University, 191 W. Woodruff Avenue, Columbus, OH
43210}
\author{C. S. Kochanek}
\affiliation{Department of Astronomy, The Ohio State University, 140 West 18th Avenue, Columbus, OH 43210, USA}
\affiliation{Center for Cosmology and Astroparticle Physics, The Ohio
State University, 191 W. Woodruff Avenue, Columbus, OH
43210}
\author{B. J. Shappee}
\affiliation{Institute for Astronomy, University of Hawai’i, 2680 Woodlawn Drive, Honolulu, HI 96822,USA}

\author{T. W. -S. Holoien}
\affiliation{Carnegie Fellow, The Observatories of the Carnegie Institution for Science, 813 Santa Barbara St., Pasadena, CA 91101, USA}

\author{T. A. Thompson}
\affiliation{Department of Astronomy, The Ohio State University, 140 West 18th Avenue, Columbus, OH 43210, USA}
\affiliation{Center for Cosmology and Astroparticle Physics, The Ohio
State University, 191 W. Woodruff Avenue, Columbus, OH
43210}
\author{J. L. Prieto}
\affiliation{N\'ucleo de Astronom\'ia de la Facultad de Ingenier\'ia, Universidad Diego Portales, Av. Ej\'ercito 441, Santiago, Chile}
\affiliation{Millennium Institute of Astrophysics, Santiago, Chile}
\author{Subo Dong}
\affiliation{Kavli Institute for Astronomy and Astrophysics, Peking
University, Yi He Yuan Road 5, Hai Dian District, China}
\author{M. Pawlak}
\affiliation{Institute of Theoretical Physics, Faculty of Mathematics and Physics, Charles University in Prague, Czech Republic}
\author{O. Pejcha}
\affiliation{Institute of Theoretical Physics, Faculty of Mathematics and Physics, Charles University in Prague, Czech Republic}



\section{}
The All-Sky Automated Survey for SuperNovae (ASAS-SN, \citealt{2014ApJ...788...48S, 2017PASP..129j4502K}) has monitored the entire visible sky to a depth of $\sim17$ mag in the V-band since 2014. ASAS-SN data is well suited for the discovery and study of variable stars \citep{2018MNRAS.477.3145J,2018arXiv180904075S}. In \citet{2018arXiv180907329J}, we have uniformly analyzed the ASAS-SN light curves of $\sim 412,000$ known variables in the VSX catalog \citep{2006SASS...25...47W}, providing homogeneously classified samples of variable groups for further study.  

\begin{figure}[htpb!]
\begin{center}
\includegraphics[scale=0.5,angle=0]{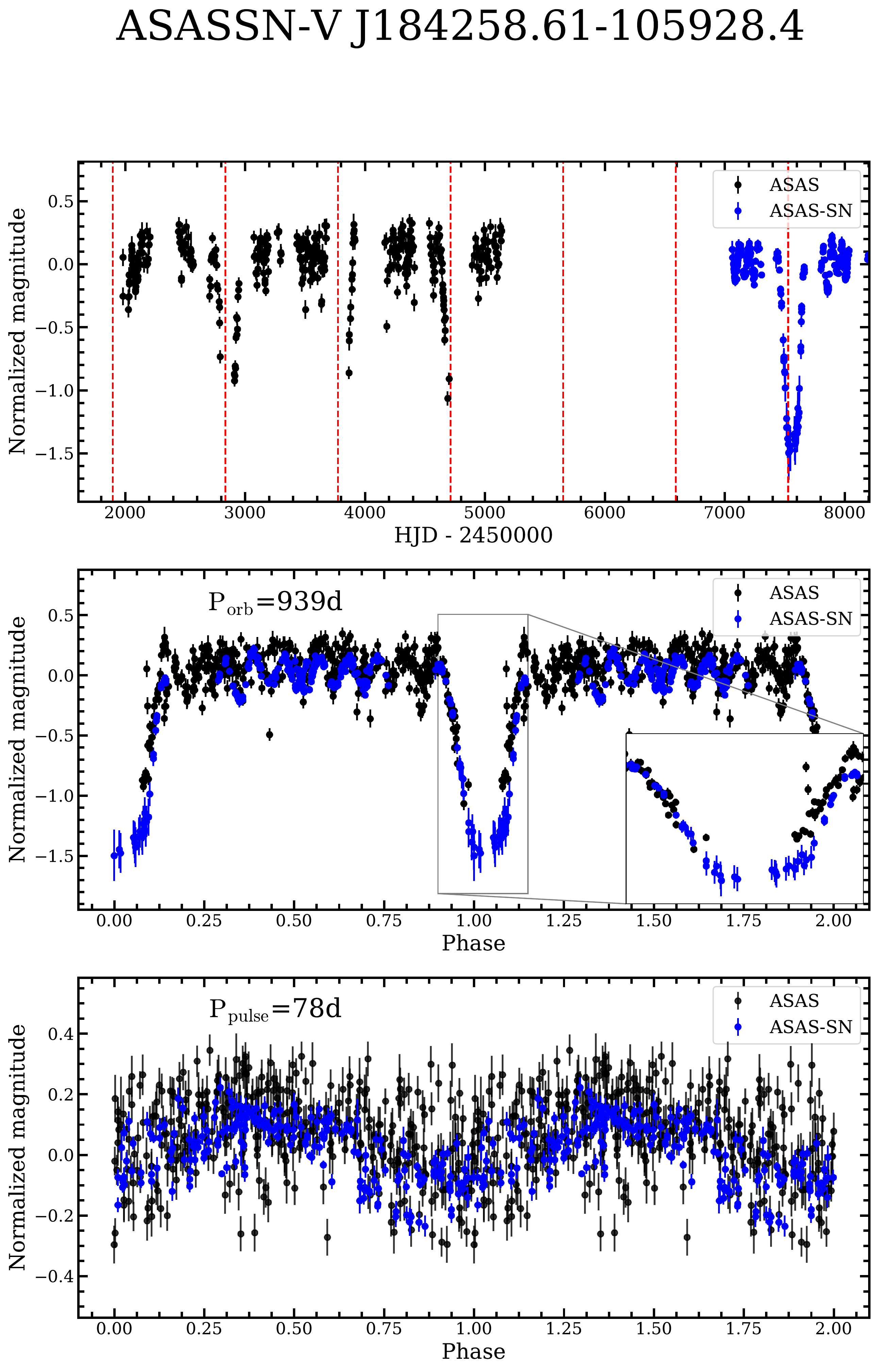}
\caption{The top panel shows the combined ASAS-SN (blue) and ASAS (black) light curve for ASASSN-V J184258.61$-$105928.4. The middle panel shows the combined light curve phased with the orbital period of $P_{\rm orb}\sim939$ d, and the bottom panel shows the combined light curve excluding the eclipses and phased with the pulsational period of $P_{\rm pulse}=78$ d.\label{fig:1}}
\end{center}
\end{figure}

Here we report the identification of a long period detached eclipsing binary system. ASASSN-V J184258.61$-$105928.4 (FY Sct) was discovered by \citet{1960AnLei..21..387H} with a spectral type of M6, and is currently classified as a semi-regular (SRB) variable in the VSX catalog. FY Sct (\textit{l,b} = $22.184$, $-3.135$) was also classified as a variable source in the INTEGRAL-OMC catalog of variable sources \citep{2012A&A...548A..79A}. The ASAS-SN data spanning 2014-2018 captures a single, deep ($A{\sim}1.5$ mag) eclipse. The similarity of this light curve to the recently identified long period detached eclipsing binary ASASSN-V J192543.72+402619.0 \citep{2018RNAAS...2c.125J} piqued our interest. We supplemented the ASAS-SN light curve with V-band data from the All-Sky Automated Survey (ASAS; \citealt{2002AcA....52..397P}), extending the light curve back to the year 2000 (Figure \ref{fig:1}). We calculated periodograms and arrived at an orbital period of $P_{\rm orb}\sim 939$ d. Based on this orbital period, we highlight the expected times of eclipse using the red dotted lines in the top panel of Figure \ref{fig:1}. The ASAS data captures three eclipses. We see clear evidence for pulsational variability when the system is out-of-eclipse, with a period of $P_{\rm pulse}=78$ d. This pulsational variability resembles the variability of semi-regular variables, and likely explains the current VSX classification as a SRB variable.

Multi-band photometry from ALLWISE \citep{2010AJ....140.1868W} and 2MASS \citep{2006AJ....131.1163S} provides colors of $W1-W2=-0.125$ and $J-K_s=1.292$ typical of an M-giant \citep{2014EAS....67..409Z}. The probabilistic Gaia DR2 distance of $d\sim 6$ kpc \citep{2018AJ....156...58B,2018arXiv180409365G} gives $M_{K_S}=-5.7$, which is also typical of a M-giant. The eclipse profile is suggestive of a disk eclipsing binary \citep{2014ApJ...788...41D} rather than a stellar companion. The primary eclipse lasts $\sim73$ days, and the next eclipse is expected to begin on ${\sim}\rm Sept.28$, 2018. We encourage further photometric and spectroscopic observations, particularly during the eclipse, to further understand this peculiar binary system.

\bibliographystyle{plainnat}

\end{document}